\documentclass[12pt]{article}
\usepackage{graphicx,color,amssymb,amsmath}
\begin{document}
\title{\textbf{The discovery of the fourth family\\at the LHC: what if?}} 

\author{B. Holdom%
\thanks{bob.holdom@utoronto.ca}\\
\emph{\small Department of Physics, University of Toronto}\\[-1ex]
\emph{\small Toronto ON Canada M5S1A7}}
\date{}
\maketitle
\begin{abstract}
The first evidence of new strong interactions may be a sufficiently massive fourth family observed at the LHC. The fourth family masses, of the leptons in particular, are constrained by the electroweak precision data, and this leads to signatures at the LHC that may imply early discovery. We study the implications of this discovery from a bottom-up perspective, where effective 4-fermion operators model the dominant effects of the new dynamics. We identify simple approximate symmetries of these operators that may be required for realistic masses of the third and fourth families. The large top mass for instance is related to the structure of these operators.
\end{abstract}

\section{The fourth family and strong interactions}
The Higgs scalar of the standard model unitarizes the scattering of massive gauge bosons, thus saving the theory from breaking down above about 1.8 TeV \cite{X}. But the Higgs by itself is not a complete solution. Additional physics must be introduced to cancel quadratically diverging contributions to the Higgs mass $m_H$, in particular from a top quark loop. The mass scale of this additional physics must be less than about $3.5m_H$ to avoid fine tuning \cite{Y}. For a light Higgs this necessitates new physics well below 1.8 TeV. Among the reasons why this type of picture is very popular are the following two. The first is that the Higgs sector and the required additional physics can all be weakly interacting, thus allowing the perturbative regime to extend to energy scales far above 1.8 TeV. The second is that the new physics that is required should be very accessible, most notably at the LHC.

On the other hand nature may have chosen a less contrived method of ensuring unitarity, one where the scale of the would-be breakdown of unitarity is the scale of new physics. This new physics would not only be responsible for electroweak symmetry breaking, but it could also be quite closely associated with the physics of flavor and fermion mass. This is in contrast to the light Higgs picture where the origin of the observed pattern of fermion masses, as encoded in a set of Yukawa couplings, is pushed to extremely high and inaccessible energies.

The theory of the Goldstone bosons of electroweak symmetry breaking may be the weak coupling description dual to a strongly coupled theory involving different degrees of freedom. In this case a simple Goldstone description only holds up to an ultraviolet cutoff, beyond which it makes more sense to use the dual description. Given that such a duality is already known to exist, relating as it does the chiral Lagrangian and quark-gluon descriptions of QCD, and given the prevalence of the duality concept in modern theoretical developments, it is curious that another manifestation of weak-strong duality is not widely anticipated to show up at the LHC.

Of course the QCD analogy for electroweak symmetry breaking has been quite well explored, as reviewed for example in \cite{C}. In technicolor theories one expects a $\rho$-like resonance to be associated with the unitarization of the Goldstone boson scattering amplitudes. A naive scaling up in mass of the QCD $\rho$ puts the new $\rho$-like state at about 2 TeV. Besides other problems with this classic technicolor picture, this broad resonance is not something that will be quickly and easily probed at the LHC. A more accessible variant is low-scale technicolor \cite{C, R}, where a new $\rho$-like state becomes both lighter and narrower. But this involves increasing the number of technifermions, and thus leads to a tension with the electroweak correction parameter $S$ that typically increases with the number of new fermions \cite{L}. For a small $S$ to emerge the theory would have to be distinctly non-QCD-like, in the sense that a constituent-quark-like approximation would have to be very poor.

We shall relax a different assumption in the original QCD analogy, namely that the new fermions involved with dynamical symmetry breaking are confined. The new fermions certainly have to feel a sufficiently attractive interaction in some channel to cause chiral symmetry breaking, but confinement is not necessary. If gauge interactions are responsible then they may be broken gauge symmetries, broken through the same dynamical fermion masses that break electroweak symmetries and/or by some other agency. This makes possible a very economical picture as far as the new fermionic degrees of freedom are concerned; a sequential fourth family with standard model quantum numbers is all that is needed.\footnote{This does not preclude the possibility that there are also other new fermions on which a new unbroken gauge symmetry continues to act. If such fermions are confined but are light or massless then their contributions to $S$ may be minimized \cite{U}. We will ignore this possibility here.} The idea that a fourth family is related to electroweak symmetry breaking has some history \cite{F0,F}.\footnote{A fourth family has also been considered for other reasons \cite{K}.}$^{,}$\footnote{For a more general review of new types of fermions see \cite{T}.}

It may seem that a further replication of the family structure, already triplicated in nature, would be the most unimaginative type of new physics that could be postulated. But a fourth family has quite profound implications if the new quarks have mass above about 550 GeV. In this case the Goldstone bosons are strongly coupled to the heavy quarks, as the classic analysis \cite{I} of partial wave unitarity shows. This precludes the perturbative description of the Goldstone modes at this energy scale and above, as would have been implied by a light Higgs. In fact the heavy quark masses would serve as the order parameters for electroweak symmetry breaking, and the new strong interactions would be expected to produce these condensates dynamically. Thus the discovery of a heavy fourth family would eliminate the raison d'\^{e}tre of both the light Higgs and the associated physics needed to protect the Higgs mass.

The discovery of a fourth family could potentially come quite early. The fourth family quarks and leptons are free to have mass mixing (CKM mixing) with the lighter fermions, and thus tree-level charged-current decays. We will discuss some processes of this type that should be quite accessible at the LHC. The only source of missing energy in these events is due to light neutrinos originating from weak interactions; this is a feature of known physics, but it is not a feature of many popular scenarios for physics beyond the standard model.

There are constraints on a fourth family. From the strong constraint on the number of light neutrinos we know that the fourth family neutrino is heavy. The $S$ parameter is sensitive to a fourth family, but the experimental limits on $S$ have been evolving over the years in such a way that the constraint on a fourth family has lessened. In addition the masses of the fourth family leptons may be such as to produce negative $S$ and $T$. As we discuss in the next section, the constraints from $S$ and $T$ do not prohibit the fourth family, but instead serve only to constrain the mass spectrum of the fourth family quarks and leptons \cite{Tneutrino,B}. The implied masses for the fourth family leptons should make them particularly accessible at the LHC, with neutrino pair production providing the most interesting signatures.

We have mentioned that the dynamical symmetry breaking of electroweak symmetries should also be quite closely associated with the physics of flavor and fermion mass. This linkage quite generally introduces some challenging issues, with the prime example being the generation of the top quark mass in a manner consistent with electroweak precision data. After the next section we shall explore such issues in the context of a heavy fourth family. Although we will not follow a top-down approach here, a sequential fourth family is theoretically attractive because it makes it possible that a theory of flavor is related to the breakdown of a simple family gauge symmetry. In contrast new fermions not having standard model quantum numbers would be more surprising and difficult to understand.

\section{Constraints and Signatures}
Constraints on the masses of the fourth family fermions $t'$, $b'$, $\tau'$ and $\nu'_{L\tau}$ are obtained from their contributions to the electroweak correction parameters $S$ and $T$. As discussed in the following sections the dynamical mass of all these fermions can arise in a similar way, including the Majorana mass for the fourth left-handed neutrino. The one loop contributions may be approximated as follows \cite{Tneutrino},
\begin{eqnarray}
S&=&\frac{7}{12\pi}-\frac{1}{3\pi}\ln(\frac{m_{\tau'}}{m_{\nu'}}),\\
\alpha f^2 T&=&\frac{1}{16\pi^2}(3g(m_{t'},m_{b'})+g(m_{\nu'},m_{\tau'}))-\frac{m_{\nu'}^2}{4\pi^2}\ln(\frac{\Lambda_{\nu'}}{m_{\nu'}}),\\
g(m_1,m_2)&=&m_1^2+m_2^2-\frac{4m_1^2m_2^2}{m_1^2-m_2^2}\ln(\frac{m_1}{m_2}),
\end{eqnarray}
where $f=246$ GeV. These expressions assume that the masses are sufficiently above the $Z$ mass; note also that $g(m_1,m_2)\rightarrow \frac{4}{3}(m_1-m_2)^2$ for $m_1\approx m_2$. The presence of an ultraviolet cutoff $\Lambda_{\nu'}$ reflects the dynamical nature of the $\nu'_\tau$ mass; namely that the mass function will fall to zero in the ultraviolet.\footnote{There is no $SU(2)_L$-triplet scalar field whose kinetic term is being renormalized by this loop, and whose vacuum expectation value would have produced a large tree-level contribution to $T$.}  We see that the lepton sector can make negative contributions to both $S$ and $T$. The Majorana nature of $\nu'_\tau$ is responsible \cite{Tneutrino} for the negative term in $T$ and the reduction of $S$ by $1/12\pi$. The origin of the mass-dependent term in $S$ is described in \cite{S}. For the values of masses that are of most interest it turns out the electroweak correction parameter $U$ is quite small, and we will ignore it henceforth. The use of these one-loop results assumes that the effects of the strong interactions are largely accounted for by using the dynamically generated masses in the loops, while ignoring momentum dependence of the masses themselves. This approximation should be more appropriate in our case of a broken gauge theory dynamics than it is for technicolor or QCD.

\begin{figure}
\begin{center}\includegraphics[%
  scale=0.42]{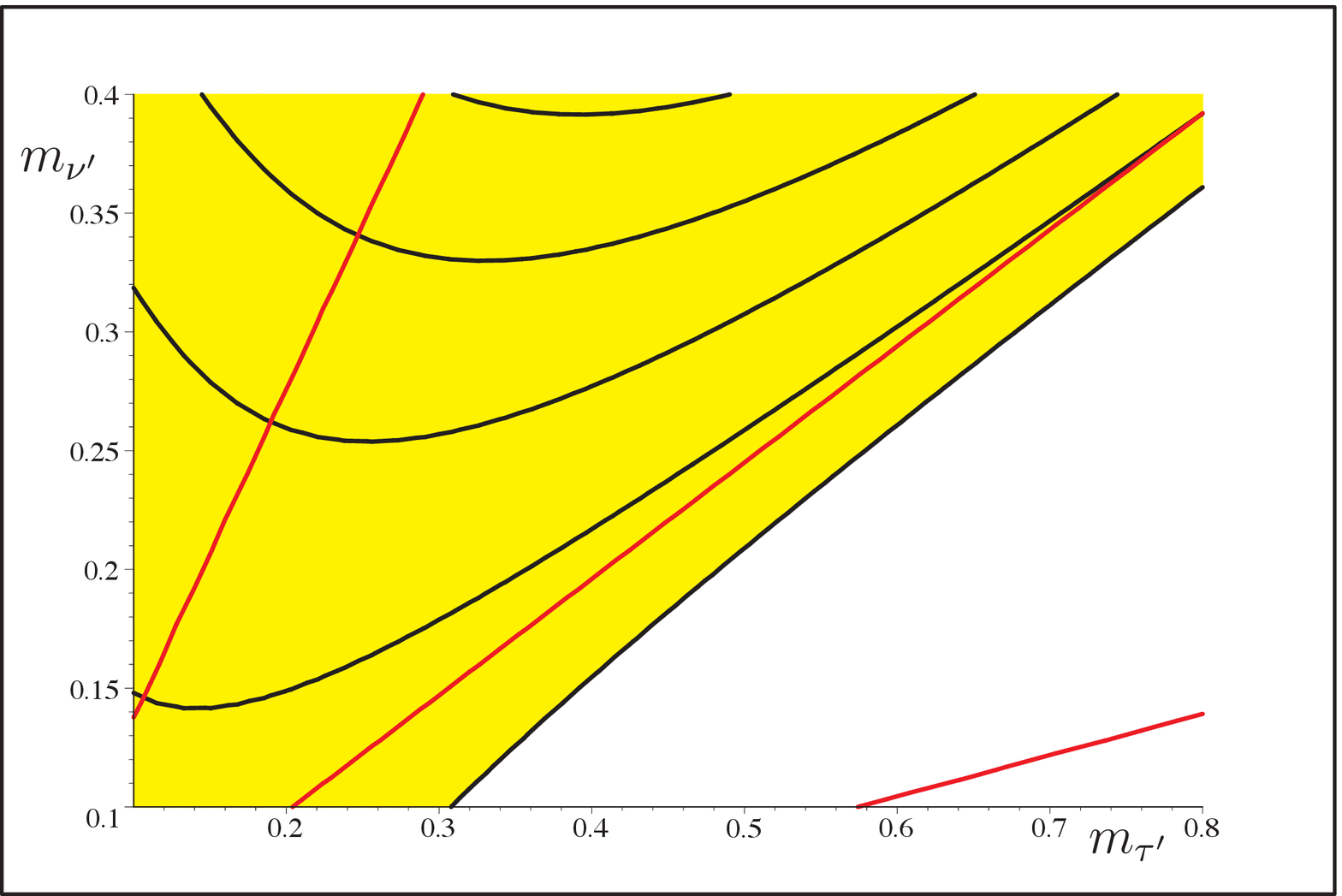}\hspace{-1pt}\includegraphics[%
  scale=0.42]{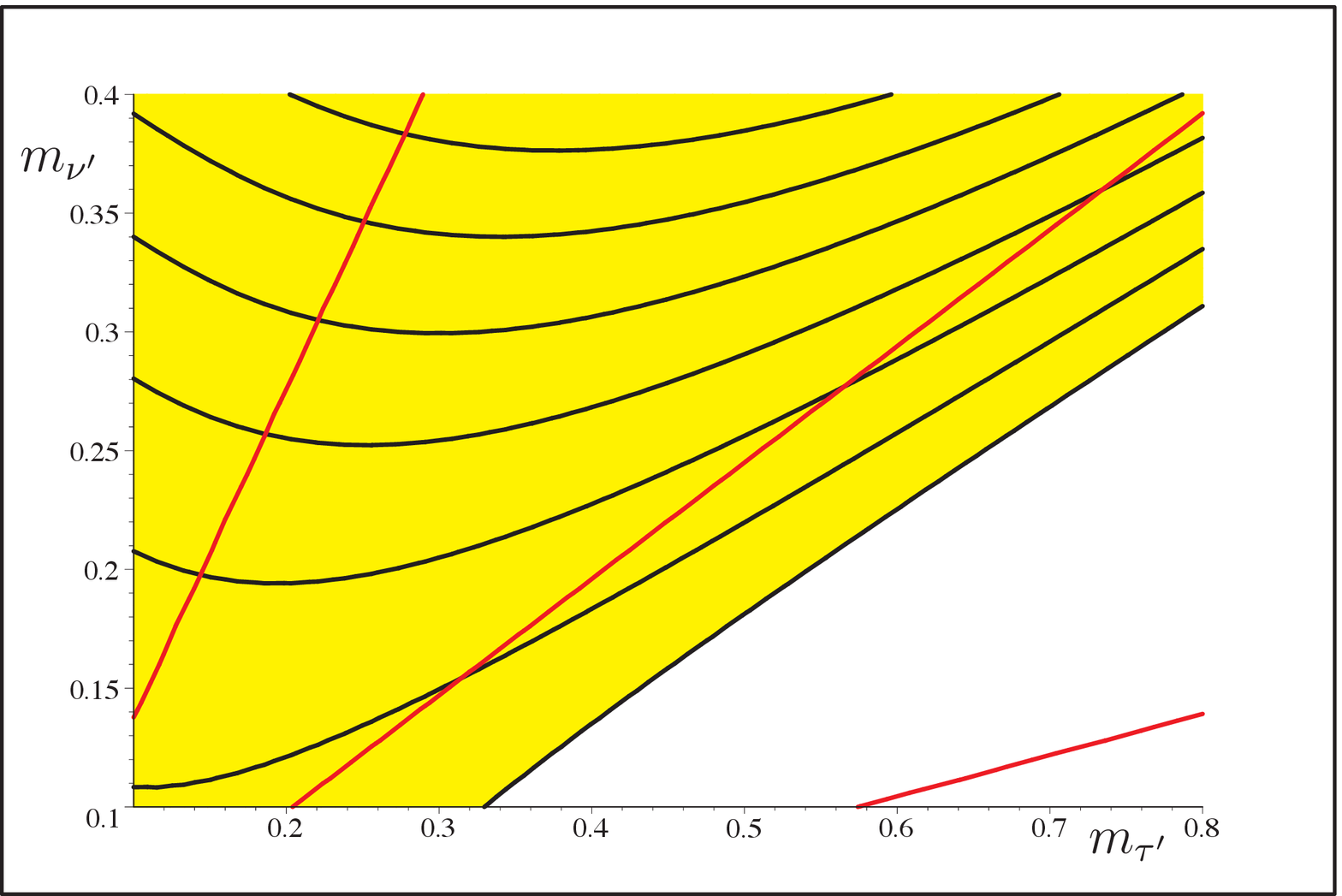}\end{center}
  \vspace{-3ex}
\caption{From the total fourth family contribution to $T$ we shade in yellow the allowed region for the $\tau'$ and $\nu'_\tau$ masses (in TeV). The successive higher contours correspond to increasing the quark contribution to $T$ by 1. The three straight red lines from bottom to top indicate when the total contribution to $S$ is $0,0.11,0.22$, where the latter two values are $1\sigma$ and $2\sigma$ away from the central measured value. The left and right figures have $\Lambda_{\nu'}=1.5m_{\nu'}$ and $2m_{\nu'}$ respectively.\label{la}}
\end{figure}

Since $T$ from the leptons can be negative, there can be some degree of cancellation between this and the positive contribution from the quarks. If we remove the light Higgs from the standard model (or set its mass to 1 TeV) then current data requires a new physics contribution to $T$ in the range $0.25\lesssim\Delta T\lesssim0.55$ at 68\% CL. (This is based on the plot at \cite{O}.)  The edge of the allowed region in the $m_{\tau'}$-$m_{\nu'}$ plane in Fig.~(\ref{la}) corresponds to lepton masses that provide the maximum contribution $\Delta T = 0.55$ along with a vanishing contribution from degenerate quarks. Within the allowed region, the leptons can provide progressively smaller and eventually negative contributions which can cancel against the progressively more positive quark contribution. Going too far into the allowed region implies more of a tuning in this cancellation, since the quark contribution to $T$ increases by one from one contour to the next.

For $S$ the constraints are such that new physics (again with the light Higgs removed) can contribute $-0.2\lesssim\Delta S\lesssim0.11$ at 68\% CL. We show the lines corresponding to the $1\sigma$ and $2\sigma$ upper bounds on $S$ along with the $S=0$ line on the plots. Thus $S$ also limits how far one can go into the allowed region. But acceptable ranges of masses remain, and this is even before realizing the uncertainties in the theoretical estimates due to strong interactions. All these considerations show that a fourth family is quite compatible with present precision data.

Taking Fig.~(\ref{la}) seriously would suggest that $m_{\tau'}/3\lesssim m_{\nu'}\lesssim m_{\tau'}/2$. We might also expect that $m_{\tau'}\lesssim m_{q'}$ due to lack of the QCD contribution to the dynamics in the lepton sector, which would tend to enhance the masses of quarks \cite{W}. It then appears plausible that $m_{\nu'}$ could be in the 150-300 GeV range, with $m_{\tau'}$ in the 400-600 GeV range. $m_{b'}$ and $m_{t'}$ may be in the 550-800 GeV range, and with a mass splitting probably not much larger than 100 GeV. Much larger mass splitting would require more tuning in the canceling contributions to $T$. But note that the total new physics contribution can be as large as half a unit of $T$, while a unit of $T$ from the quarks corresponds to about a 130 GeV quark mass splitting, so even that much splitting would not constitute a fine tuning. In section 4 we shall argue that $m_{b'}>m_{t'}$.

The first signal of a fourth family may involve the fourth family leptons. $\nu'_\tau\overline{\nu}'_{\tau}$ production is more interesting than $\overline{\tau}'\tau'$ production, both because $\nu'_\tau$ is expected to be lighter and because its decay modes are more interesting. The decay $\nu'_\tau\rightarrow\ell W$ with $\ell=(\tau\mbox{ or }\mu\mbox{ or }e)$ leads to the following final states from $\nu'_\tau\overline{\nu}'_{\tau}$ production,
\begin{equation}
2\ell+4j,\quad\quad3\ell+2j+E\!\!\!\!/,\quad\quad4\ell+E\!\!\!\!/.
\end{equation}
These processes should be quite accessible at the LHC, although serious studies seem to be lacking. The first process can have same-sign leptons due to the Majorana nature of the $\nu'_\tau$. (This and other properties of Majorana neutrino pair production are discussed in \cite{G}.) The last process may be similar to the production and decay of a pair of neutralinos, but the presence of the other two processes should make the distinction between neutrinos and neutralinos clear.

The expected heavy quark decays are $t'\rightarrow bW$, which would look like a heavy $t$ decay, and $b'\rightarrow tW$. But if the associated CKM mixing is small then $b'\rightarrow t'W$ could dominate $b'\rightarrow tW$ if the $b'$-$t'$ mass splitting is large enough. Even if the $W$ has to be virtual due to a small mass difference (thus implying phase space suppression) the $b'\rightarrow t'W^{(*)}$ process could still be significant when the mixing is small enough. Thus a process of interest is $pp\rightarrow b'\overline{b}'\rightarrow t'\overline{t}'WW \rightarrow b\overline{b}WWWW$. Notice that the $b$ jets can be particularly hard and isolated, and appropriate cuts can help to reduce the background from $t\overline{t}$ production. This has been used in a study of the $pp\rightarrow t'\overline{t}'\rightarrow b\overline{b}WW$ process at the LHC \cite{A}. The $b'\overline{b}'$ process has two extra $W$'s, aiding further the discrimination from background. One of the resulting signals involves two same-sign leptons and missing energy along with the jets.\footnote{Note that if $t'$ is in fact the heaviest, then the process $pp\rightarrow t'\overline{t}'\rightarrow b'\overline{b}'WW \rightarrow t\overline{t}WWWW \rightarrow b\overline{b}WWWWWW$ is possible.}

\section{Flavor Physics}
Starting with a massless gauge theory of fermions, we suppose that mass and flavor emerges through the breakdown of some of the gauge symmetries. At scales 100 to 1000 TeV some interactions are most likely both strong and chiral, and we assume that they lead in some economical manner to their self-breaking at these scales. The effects of this flavor physics dynamics on lower scales will be carried by a set of effective operators. We expect that all possible operators allowed by the unbroken symmetries are generated, even those that can only be generated nonperturbatively. These manifestations of nonperturbative physics will be important in the following.  The only masses allowed by the unbroken SU(2)$_L\times$U(1) symmetry are right-handed neutrino masses; all other fermions are protected from receiving a flavor scale mass and at lower scales will only be affected by the flavor physics through multi-fermion and other nonrenormalizable operators. 

The mass of the top quark will certainly be well within an order of magnitude of the $t',b'$ masses, and this suggests that the physics origin of these three masses should be somehow related. We will take this as a strong hint to consider the possibility that the third family also experiences 4-fermion interactions of the same form and similar magnitude as the interactions involving the fourth family. This leads to the picture where the original flavor gauge symmetry breaks in such a way that the first two families are singlets under an unbroken remnant. This remnant gauge symmetry acts on the third and fourth families and may only break closer to the TeV scale. It will contribute to the anomalous scaling of the various operators, and it may ensure that certain operators remain significant at the TeV scale, even though they are generated at the flavor scale. In particular we assume that the theory exhibits near conformal scaling for some range of scales above a TeV, in which case $\overline{\psi}\psi$ has an effective scaling dimension close to 2 \cite{P}. This makes natural the possibility that some 4-fermion operators, at least those that are composed of two such scalars, are close to being relevant operators (close to scaling dimension 4). The role of enhanced operators of this form in theories of flavor has been noted before \cite{F0,Q,tmass,D}. In the following we shall focus on operators of the scalar-scalar form and composed of third and fourth family fermions.

We notice how the same fermions, four standard model families, remain the fundamental degrees of freedom throughout the range of energy scales, even though they experience strong interactions at various scales. The light fermions only feel the strong interactions at the flavor scale, while the heavy families also feel strong interactions down to the TeV scale. These latter interactions become strong enough for the fourth family masses to form at the TeV scale. And even then, since the fermions do not become confined, it is still useful to describe the physics of interest at the LHC in terms of the massive fermion degrees of freedom. We note that a massive constituent quark description works quite well in QCD, even though the quarks in that case are confined. The massive quark picture should be even more appropriate in our case.

We are thus led to a phenomenological description of the dynamics responsible for a condensate $\langle \overline{q'}q'\rangle$ of the fourth family quarks $q'\equiv (t',b')$. The Nambu-Jona-Lasinio (NJL) model provides a minimal framework, where this dynamics is described by a 4-fermion interaction,
\begin{equation}
\frac{g^2}{\Lambda^2}(\overline{q}'_{L} q'_{R})(\overline{q}'_{R} q'_{L})
.\label{e1}\end{equation}
$\Lambda$ represents a cutoff above which a softening of this interaction should occur in a more realistic description. For $g$ above some critical value $g_c$ a condensation occurs. Without invoking a fine-tuning of $g$ close to $g_c$, the resulting dynamical mass $m_{q'}$ should not be too far below $\Lambda$. To get a sense of the fine tuning needed for $m_{q'}\ll\Lambda$, we note that a light composite scalar emerges with mass $\approx 2m_{q'}$ in this case \cite{N}. Then contributions of order $\Lambda^2$ to the scalar mass must be fine-tuned away, and thus the degree of fine-tuning is $\approx 4m_{q'}^2/\Lambda^2$. We believe that fine tuning does not naturally occur and that $m_{q'}$ is not much below $\Lambda/2$.

There is a relation between $m_{q'}$ and $\Lambda$ and the electroweak symmetry breaking scale $v=246$ GeV which is given in a one-loop approximation by the Pagels-Stokar formula
\begin{equation}
v^2=f^2\approx\frac{3m_{q'}^2}{4\pi^2} \ln\frac{\Lambda^2}{m_{q'}^2}
.\end{equation}
For example for $m_{q'}\approx 750$ GeV and $\Lambda$ roughly twice that would imply a suitable $v$ from this formula.\footnote{This is assuming that the $q'$ quarks give the dominant contribution to $v$; the additional smaller contribution from the fourth family leptons implies a somewhat smaller $m_{q'}$.} But ambiguities in matching the phenomenological NJL model to the underlying theory implies that $m_{q'}$ as low as 500 GeV may also be acceptable. This is in line with the unitarity analysis \cite{I}.

In the next two sections we wish to explore the naturalness of finding the third and fourth family masses emerging in this type of picture. Our task will be to understand not only the origin of top mass, but also the smaller masses of the other members of the third family. Rather than trying to specify more precisely what the flavor interactions are, as in \cite{D}, we will continue with a bottom-up approach, and try to find a minimal set of constraints on the 4-fermion operators that could allow for realistic masses. Constraints that can be expressed in terms of approximate symmetries have some chance of being realizable by some underlying flavor dynamics.

\section{Approximate Symmetries}
There are two anomaly-free U(1) family symmetries of the third and fourth families that we could consider. The generators have charges $(+,+,-,-)$ and $(+,-,-,+)$  for the fields $(\psi'_L,\psi'_R,\psi_L,\psi_R)$, where $\psi$ and $\psi'$ denote members of the third and fourth families respectively. They are chosen so that they are vector-like and axial-like respectively with respect to the fermion mass eigenstates, and either or both may correspond to gauge symmetries of the high scale flavor physics. Both symmetries must be broken. Of the two, the axial one is of more interest for constraining the operators that are relevant for producing masses; we will label it by $\cal{Q}$. We no longer consider the possible vector-like symmetry. Notice that $\cal{Q}$ is broken at the very least by the $\langle \overline{q'}q'\rangle$ condensate, and if there is no other much larger contribution to its breaking then it will be a useful approximate symmetry to constrain operators. In particular it will help us to understand the $b$ to $q'$ mass ratio.

We can also consider another axial charge, $(+,-,+,-)$, labeled by $\tilde{\cal{Q}}$. This is not anomaly-free and so could not be gauged, and we take it to be a more badly broken symmetry than $\cal{Q}$. The operators that respect $\tilde{\cal{Q}}$ include those that can be generated by gauge boson exchange diagrams, while those that violate $\tilde{\cal{Q}}$ are purely nonperturbative. Since the two classes of operators are generated by distinctly different physics it is not unnatural to assume that $\tilde{\cal{Q}}$-violating operators are somewhat suppressed relative to the $\tilde{\cal{Q}}$-invariant ones. This suppression will give rise to the $t$ to $q'$ mass ratio. 

The quark operators we consider are the products of the following color-singlet, Lorentz scalars, where the products are constructed to preserve SU(2)$_L\times$U(1).
\begin{eqnarray}
&&\overline{q}'_Lq'_R\quad\overline{q}'_Lq_R\quad\overline{q}_Lq'_R\quad\overline{q}_Lq_R\nonumber\\
&&\overline{q}'_Rq'_L\quad\overline{q}'_Rq_L\quad\overline{q}_Rq'_L\quad\overline{q}_Rq_L
\label{e2}\end{eqnarray}
These scalars are either $\cal{Q}$-charged or $\cal{Q}$-neutral, but we only consider products that are $\cal{Q}$-invariant. The product of a scalar in the top row with a scalar in the second row produces a $\tilde{\cal{Q}}$-invariant 4-fermion operator,  with the LRRL structure as in (\ref{e1}). The product of two operators within a row produces a $\tilde{\cal{Q}}$-violating operator with the LRLR structure, for example
\begin{equation}
\label{e7}\epsilon_{ij}\epsilon_{kl}\overline{q}'_{Li}q'_{Rk}\overline{q}_{Lj}q_{Rl},
\end{equation}
which is SU(2)$_L\times$SU(2)$_R$ invariant. But as we shall see, SU(2)$_R$ violation must manifest itself in the $\tilde{\cal{Q}}$-violating operators, reflecting the SU(2)$_R$ breaking that must originate in the associated nonperturbative dynamics.

Depending on the signs and strengths of all these interactions we assume that condensates form. It is then a question of vacuum alignment as to whether the $\cal{Q}$-charged or $\cal{Q}$-neutral condensates form. We have already assumed the former; more precisely we have assumed that some approximate symmetry exists, labelled by $\cal{Q}$, which is axial with respect to the mass eigenstate basis.

The dynamics that produces $\cal{Q}$-charged condensates is represented by the 4-fermion operators that involve the $\cal{Q}$-charged scalars. There are only two such operators that are both $\cal{Q}$-invariant and $\tilde{\cal{Q}}$-invariant, 
\begin{equation}
\overline{q}'_Lq'_R\overline{q}'_Rq'_L,\quad\quad\overline{q}_Lq_R\overline{q}_Rq_L
.\label{e5}\end{equation}
It is important to note that $\overline{q}'_Lq'_R\overline{q}_Rq_L$ is not $\cal{Q}$-invariant. Although these two operators may have similar (running) coefficients we assume (in the absence of a symmetry) that they are not identical. Then we can assume that the first operator develops an effective coupling above the critical value, while the second operator does not. Alternatively or in addition there may be an effective coupling between the two channels that discourages both condensates from forming simultaneously. This type of coupling between channels could be represented by the multi-quark operator $\overline{q}'_Lq'_R\overline{q}'_Rq'_L\overline{q}_Lq_R\overline{q}_Rq_L$ with the appropriate sign.

Thus if these $\tilde{\cal{Q}}$-invariant operators respect SU(2)$_R$, and if we continue to ignore the $\tilde{\cal{Q}}$-violating operators we can have the result,
\begin{eqnarray}
&&\langle\overline{t}'t'\rangle=\langle\overline{b}' b'\rangle\neq0,\\
&&\langle\overline{t} t\rangle=\langle\overline{b} b\rangle=0.
\end{eqnarray}
Then to obtain a $t$ mass from a $\cal{Q}$-invariant operator we must turn to $\tilde{\cal{Q}}$-violating operators. The operator of interest is
\begin{equation}
\epsilon_{ij}\overline{q}'_{Li} b'_R\overline{q}_{Lj} t_R\rightarrow\overline{b}'_L b'_R\overline{t}_L t_R
.\label{e3}\end{equation}
This type of operator must involve both families to be $\cal{Q}$-invariant, and we see that it feeds mass from $b'$ to $t$. Thus we see that the $t$ to $b'$ mass ratio is a measure of the amount of $\tilde{\cal{Q}}$ violation.

There is a corresponding operator that feeds mass from $t'$ to $b$, and thus that operator must be significantly smaller.  The dominance of the operator in (\ref{e3}) indicates that there must be a close to maximal breakdown of SU(2)$_R$ in the $\tilde{\cal{Q}}$-violating sector of the underlying dynamics.\footnote{A toy scalar potential was considered in the appendix of the second reference in \cite{D} that illustrates such a maximal breakdown of SU(2)$_R$.} The $b$ mass could also be produced by the $\tilde{\cal{Q}}$-invariant but $\cal{Q}$-violating operator $\overline{b}'_L b'_R\overline{b}_R b_L$. Thus the $b$ to $b'$ mass ratio puts an upper bound on the amount of $\cal{Q}$ violation in the quark sector.

Obtaining a large enough $t$ mass has often proven to be difficult in models of dynamical symmetry breaking.  This is because the operator responsible for the $t$ mass has typically been taken to be generated by a simple gauge boson exchange. In our context this would correspond to the $\tilde{\cal{Q}}$-invariant operator $\overline{q}_L\gamma_\mu q'_L\overline{t}'_R\gamma^\mu t_R\rightarrow\overline{t}'_L t'_R\overline{t}_R t_L$. The trouble is that if this operator was generated by the exchange of a relatively light gauge boson (it cannot be in our context because it is both $\cal{Q}$ and SU(2)$_R$ violating) then the following operators could also be generated through closely related gauge boson exchanges:
\begin{equation}
\overline{q}'_L\gamma_\mu q'_L\overline{t}'_R\gamma^\mu t'_R,\quad\quad\overline{q}_L\gamma_\mu q'_L\overline{q}'_L\gamma^\mu q_L
.\label{e4}\end{equation}
The first of these operators would give rise to a mass splitting in the $(t',b')$ doublet of the same order as the top mass itself. A splitting equal to the top mass produces a shift $\Delta T\approx1.7$, which is significantly larger than what is currently allowed. For a more detailed analysis of this problem in the technicolor context see \cite{V}. The second operator implies a correction to the $Zb\overline{b}$ vertex that is similarly too large \cite{M}.

These basic problems have motivated many different types of model building efforts such as non-commuting extended technicolor, multiscale technicolor, topcolor, topcolor-assisted technicolor and topcolor seesaw models (for a review and references see \cite{C}).\footnote{The same problems also require special attention in the Higgless models of higher dimensions \cite{J}.} These models generally involve complicating the gauge structure and/or adding new gauge dynamics coupling to the $t$ quark. Here we are pointing out that it is not strictly necessary to invoke such complications, given the possibility that the operator (\ref{e3}) gives the dominant contribution to the $t$ mass. 

The point is that the side-effects of the $\tilde{\cal{Q}}$-violating operator (\ref{e3}) are not so severe \cite{tmass}. It can give rise to effects similar to those in (\ref{e4}) (which are $\tilde{\cal{Q}}$-invariant) only by inserting it twice in a loop. Thus an operator similar to the first operator of (\ref{e4}), $\overline{q}'_L b'_R\overline{b}'_R q'_L$, is generated with a suppression of $m_t^2/m_{q'}^2$ along with the extra loop suppression. This effect breaks the SU(2)$_R$ invariance of the operators $\overline{q}'_L q'_R\overline{q}'_R q'_L$ that are responsible for the $t'$ and $b'$ masses, giving rise to a mass splitting with $m_{b'}>m_{t'}$. The contribution to $T$, proportional to $(m_{b'}-m_{t'})^2$, is then suppressed at least by $m_t^4/m_{q'}^4$ in comparison to the quadratic suppression in models with only gauge-exchange operators.\footnote{The operator $\overline{b}'_R\gamma_\mu b'_R\overline{b}'_R\gamma^\mu b'_R$ that can contribute directly to $T$ would require four insertions of operator (\ref{e3}) and three loops.} An operator that can affect the $Zb\overline{b}$ vertex, like the second operator in (\ref{e4}), but which can only be generated by a loop with two insertions of operator (\ref{e3}) is $\overline{q}_L b'_R\overline{b}'_R q_L$.\footnote{The SU(2)$_R$ invariant operators in (\ref{e5}), and those closely related to them such as $\overline{q}_{L} q'_{R}\overline{q}'_{R} q_{L}$, neither contribute to $T$ nor correct the $Zb\overline{b}$ vertex.} (The second operator in (\ref{e4}) is not generated.) In conclusion we see how the corrections to $T$ and the $Zb\overline{b}$ vertex are more shielded from top mass generation because of the $\tilde{\cal{Q}}$-violating nature of the top mass operator.

\section{Leptons}
We first turn to the charged lepton sector. For $\tau$ and $\tau'$ we can suppose similar 4-fermion dynamics as in the quark sector, with the same approximate $\cal{Q}$ symmetry constraining the dynamics. Thus we can again suppose that the $\cal{Q}$ and $\tilde{\cal{Q}}$ invariant operators (the analogs of (\ref{e5}) with $\tau$ and $\tau'$ replacing $q$ and $q'$) generate $\langle\overline{\tau}'\tau'\rangle\neq0$ while $\langle\overline{\tau}\tau\rangle=0$. The $\tau$ mass can arise similarly to the $b$ mass, and in particular the following SU(2)$_R$ and $\tilde{\cal{Q}}$ violating, but $\cal{Q}$-invariant operators can feed mass from $t'$ to $b$ and $\tau$:
\begin{equation}
\overline{t}'_L t'_R\overline{b}_L b_R,\quad\quad\overline{t}'_L t'_R\overline{\tau}_L \tau_R
.\label{e13}\end{equation}
Here we see our first instance of an operator with both quarks and leptons. (In the Appendix we consider a different choice of the approximate symmetries that results in a different structure for the mixed operators.)

Neutrinos are more special. We are supposing that all fermions, including the right-handed neutrinos, participate in the strong flavor interactions at the flavor scale. If SU(2)$_L\times$U(1) is the only exact chiral symmetry remaining below the flavor scale, then there is nothing to protect the right-handed neutrinos from receiving mass from the strong interactions. In fact right-handed neutrino condensates serve as excellent order parameters not only for the breakdown of flavor symmetries, but also for the breakdown of enlargements of the electroweak symmetry such as those involving SU(2)$_R\times$U(1)$_{B-L}$ and/or Pati-Salam-like gauge interactions. With their masses at the flavor scale the right-handed neutrinos are absent in the theory below the flavor scale, and this in turn is important for understanding why the small left-handed neutrino masses are so dramatically different from other fermion masses.

But first we consider $\nu'_{L\tau}$ where we see that its mass (again $\cal{Q}$-violating) can arise in a similar way to other fourth family members. Again there are only two $\cal{Q}$ and $\tilde{\cal{Q}}$ invariant operators of interest,
\begin{equation}
\ell'_L\ell'_L({\ell'_L}{\ell'_L})^\dagger,\quad\quad\ell_L\ell_L({\ell_L}{\ell_L})^\dagger
,\label{e9}\end{equation}
since $\ell'_L\ell'_L({\ell_L}{\ell_L})^\dagger$ is not $\cal{Q}$-invariant. (Operators such as $\ell'_L\ell'_L{\ell_L}{\ell_L}$ can be $\cal{Q}$ and SU(2)$_L\times$U(1) invariant, but they don't involve four neutrinos.) Thus by the same reasoning as before we can assume that $\langle\nu'_{L\tau}\nu'_{L\tau}\rangle\neq0$ while $\langle\nu_{L\tau}\nu_{L\tau}\rangle=0$. We are then left with the three light neutrinos $(\nu_{L\tau},\nu_{L\mu},\nu_{Le})$.

Now the question is whether $\nu_{L\tau}$ can receive a mass in a manner similar to other third family fermions. The answer is no, since in this case there are no $\cal{Q}$-invariant operators that can feed down mass from the fourth family. The $\cal{Q}$-violating operator $\ell'_L\ell'_L({\ell_L}{\ell_L})^\dagger$ can yield a $\nu_{L\tau}$ mass, and thus the relatively tiny value of this mass implies that the $\cal{Q}$ symmetry must be very well preserved by the effective operators in the left-handed lepton sector.

There are also operators that arise by integrating out the right-handed neutrinos at the flavor scale. The resulting lepton number violating operators necessarily involve six fermions, and they can generate Majorana masses for $\nu_{L\mu}$ and $\nu_{Le}$ as well as $\nu_{L\tau}$. These 6-fermion operators are naively suppressed by three more powers of the flavor scale compared to 4-fermion operators, thus providing a natural mechanism for the suppression of neutrino masses. This could be thought of as a type of see-saw mechanism, but the right-handed neutrino mass in the see-saw is now set by the flavor scale, of order 1000 TeV. Once again we see how the absence of a Higgs brings down a mass scale of interest.

There are many different 6-fermion operators that can contribute. If they are to feed down mass from the heaviest fermions then they can be constructed by taking Lorentz invariant products of any pair of the following 3-fermion operators (all of which transform as SU(2)$_L\times$U(1) invariant Lorentz spinors).
\begin{equation}
\overline{t}'_L t'_R \nu_{Li}\quad\quad\overline{t}_L t_R \nu_{Li}\quad\quad\overline{b}'_R b'_L \nu_{Li}\quad\quad\overline{\tau}'_R \tau'_L \nu_{Li}\quad\quad i=e,\mu,\tau
\end{equation}
We see that each element of the $3\times3$ Majorana neutrino mass matrix has many possible contributions from the various combinations. The relative size of these contributions depends on the detailed structure of the flavor interactions and their breakdown. By dimensional analysis the resulting neutrino masses are probably no less than $(600 \mbox{ GeV})^6/(1000 \mbox{ TeV})^5\approx5\times10^{-5} \mbox{ eV}$. This is likely an underestimate since it ignores possible anomalous scaling enhancement of the 6-fermion operators. One is also tempted to use the see-saw estimate of the form $m^2/M$, where $m$ is some Dirac mass and $M$ is the right-handed neutrino mass, but this assumes that the anomalous scaling contained in the value of $m^2$ is the same as that of the 6-fermion operator. This is certainly incorrect for the case of $\nu_{L\tau}$ but it may be more appropriate for $\nu_{L\mu}$ and $\nu_{Le}$. Reasonable masses seem entirely possible (for example if $m^2\approx m_e m_\mu$). In addition we see that the structure of the $3\times3$ neutrino mass matrix is quite unrelated to the quark and charged lepton mass matrices, and can have significant off-diagonal terms and thus large mixings \cite{D}.

\section{Further remarks}
We return to the question of the CKM mixing in the quark sector, responsible for the decays $t'\rightarrow bW$ and $b'\rightarrow tW$. The off-diagonal $\overline{t}t'$ or $\overline{b}b'$ mass elements would require $\cal{Q}$ violation, thus making this CKM mixing naturally small. Alternatively these off-diagonal elements could arise as described in the Appendix. As another possibility, \cite{CP} shows that kinetic-term mixing effects may be a source of CKM mixing along with CP violating phases. Flavor physics could also generate flavor changing neutral current decay modes of the heavy quarks \cite{H}. But these vertex-type mixing effects are probably smaller than the mass mixing effects, due to less anomalous scaling enhancement of the relevant operators, and thus we expect the charged current decays to dominate.

Pair production of the fourth family fermions could exhibit a resonance structure associated with the physics near the cutoff of our effective theory. For example there could be a broken U(1) gauge boson that mixes with the $Z$ and which couples strongly to the fourth (and third) families. Alternatively the strong interactions may imply unconfined bound states of the heavy fermions. And finally if the CKM mixing is small enough then even QCD bound states of the heavy quarks could show up as resonances.

There may also be approximate global symmetries that are broken by the fourth family condensates leading to pseudo-Goldstone bosons, similar to technipions of technicolor theories and coupling to fermions in similar ways. But the masses of such states are so extremely model dependent that we consider them no further. We note though that our practice of assuming the existence of all possible multi-fermion operators generally eliminates the concern over unwanted light or massless pseudo-Goldstone bosons, especially if the original underlying theory has no global symmetries to begin with.

We have been concerned with a $\cal{Q}$ invariance of operators involving only the third and fourth family fermions. This is only an approximate symmetry of flavor physics in particular because, if the light fermions are $\cal{Q}$ neutral, it cannot be a symmetry of operators that are needed to feed mass to the light fermions. It may be possible to extend the $\cal{Q}$ generator to also act on light fermions and thus find an approximate symmetry of a larger set of operators and the full mass matrices. This would lead to the consideration of more complete models, where the full particle content of the theory and the assumed pattern of symmetry breaking are both specified. Such a top-down approach was taken in \cite{D}, and there it may be seen that the $\cal{Q}$ generator and its extension to the light families is a gauge generator of the complete underlying theory. One comment about such a picture is that the hierarchy between the third and fourth family masses may lead in turn to the hierarchy between the first two families. We have chosen in this work to focus in a more model independent fashion on the heavy families, since this is where the more serious issues typically arise.

In summary, a sufficiently massive fourth family points towards an extension of the standard model that treats the Goldstone bosons of electroweak symmetry breaking as the weak coupling dual description of a more fundamental strongly coupled theory. Although we have not specified the fundamental interactions of the fourth (and third) families, we have modeled them phenomenologically via 4-fermion operators. This has enabled us to find some minimal approximate U(1) symmetries of the fundamental interactions that help to explain the range of masses of the third and fourth families. This makes it more likely that such interactions can exist. 

The fourth family forms part of the fundamental degrees of freedom, and it may constitute all of the new fermionic degrees of freedom. The fourth family quark masses are fixed (up to theoretical uncertainties) by the scale of electroweak symmetry breaking, and then the masses of the fourth family leptons are constrained by the $T$ (and $S$) parameters. This is analogous to the Higgs picture where the vacuum expectation value $v$ is fixed and there is another parameter, the Higgs mass, that must be adjusted small enough to obtain the correct $T$. Additional new physics is required to protect the Higgs mass. It is exciting to realize that within a few years we will know which picture of new physics comes closer to describing reality.

\section*{Appendix: Alternative choice of approximate symmetries}
We have seen how quark masses can affect the lepton mass matrix, and vice versa, but the structure of these mixed operators may be different than described above. To see this we reconsider the possible anomaly free symmetries of the third and fourth families, now generalizing to generators that do not act identically on quarks and leptons.
\begin{center}
\begin{tabular}{ccccccccc} & $q'_L$ & $q'_R$ & $\ell'_L$ & $\ell'_R$ & $q_L$ & $q_R$ & $\ell_L$ & $\ell_R$ \\
${\cal Q}_V^q$ & $+$ & $+$ & 0 & 0 & $-$ & $-$ & 0 & 0 \\
${\cal Q}_A^q$ & $+$ & $-$ & 0 & 0 & $-$ & $+$ & 0 & 0  \\
${\cal Q}_V^\ell$ & 0 & 0 & $+$ & $+$ & 0 & 0 & $-$ & $-$ \\
${\cal Q}_A^\ell$ & 0 & 0 & $+$ & $-$ & 0 & 0 & $-$ & $+$ \end{tabular}
\end{center}
These cannot all be independent approximate symmetries, since that would suppress any mixed operator, such as the second operator in (\ref{e13}). Thus far we have only needed to assume that ${\cal Q}_A^q+{\cal Q}_A^\ell$ (which we labeled simply as $\cal Q$) is an approximate symmetry. But an interesting alternative is to assume that the following two are approximate symmetries: ${\cal Q}_A^q+{\cal Q}_V^\ell$ and ${\cal Q}_V^q+{\cal Q}_A^\ell$. The effect on the pure quark or pure lepton operators of interest to mass formation would be the same as before. But the mixed operator in (\ref{e13}) would not be allowed, and instead there could be the following operators:
\begin{equation*}
\overline{t}'_L t'_R\overline{\tau}_L \tau'_R,\quad\quad\overline{b}'_L b'_R\overline{\tau}_R \tau'_L.
\end{equation*}
This would give rise to off-diagonal mass elements in the charged lepton mass matrix, which along with the $\tau'$ mass would produce a $\tau$ mass in a see-saw manner. Similarly there could be new off-diagonal elements in the quark mass matrix, for example from the operator $\overline{\tau}'_L \tau'_R\overline{t}_L t'_R$, thus creating new sources of CKM mixing \cite{D}.

\section*{Acknowledgments}
This work was supported in part by the National Science and Engineering Research Council of Canada.


\begin{thebibliography}{11}
\bibitem{X} M.~Chanowitz and M.~K.~Gaillard, Nucl.~Phys.~B261:379, 1985.
\bibitem{Y} B.~Barbieri and A.~Strumia, ``The LEP Paradox'', hep-ph/0007265.
\bibitem{C} C.~T.~Hill, E.~H.~Simmons, Phys.~Rept.~381:235, 2003, Erratum-ibid.390:553, 2004, hep-ph/0203079.
\bibitem{R} K.~Lane, ``Search for Low-Scale Technicolor at the Tevatron'', hep-ph/0605119, and references therein.
\bibitem{L} B.~Holdom and J.~Terning Phys.~Lett.~B247:88, 1990; M.~Peskin and T.~Takeuchi Phys.~Rev.~Lett.~65:964, 1990; M.~Golden and L.~Randall Nucl.~Phys.~B361:3, 1991.
\bibitem{U} B.~Holdom, Phys.~Lett.~B314:89, 1993.
\bibitem{F0} B.~Holdom, Phys.~Rev.~Lett.~57:2496, 1986, Erratum-ibid.58:177, 1987 (corrects the many random errata introduced by the journal); B.~Holdom, Phys.~Lett.~B143:227, 1984; B.~Holdom, Phys.~Lett.~B246: 169, 1990.
\bibitem{F} C.~T.~Hill, M.~A.~Luty, E.~A.~Paschos, Phys.~Rev.~D43:3011, 1991; T.~Elliott and S.~F.~King, Physics Letters B283:371, 1992; C.~T.~Hill, D.~C.~Kennedy, T.~Onogi, H.-L.~Yu, Phys.~Rev.~D47:2940, 1993.
\bibitem{K} A.~Datta and S.~Raychaudhuri, Phys.~Rev.~D49:4762, 1994;
A.~Celikel, A.~K.~Ciftci and S.~Sultansoy, Phys.~Lett.~B342:257, 1995;
\bibitem{T} P.~H.~Frampton, P.~Q.~Hung, M.~Sher, Phys.~Rept.~330:263, 2000.
\bibitem{I} M.~S.~Chanowitz, M.~A.~Frurman and I.~Hinchliffe, Nuclear Physics B153:402, 1979.
\bibitem{Tneutrino} B.~Holdom, Phys.~Rev.~D54:721, 1996, hep-ph/9602248.
\bibitem{B} M.~Maltoni, V.~A.~Novikov, L.~B.~Okun, A.~N.~Rozanov, and M.~I.~Vysotsky, Phys.~Lett.~B476:107, 2000, hep-ph/9911535; H.-J.~He, N.~Polonsky, S.~Su, Phys.~Rev.~D64:053004, 2001, hep-ph/0102144.
\bibitem{S} E.~Gates and J.~Terning, Phys.~Rev.~Lett.~67:1840, 1991; T.~Appelquist and J.~Terning, Phys.~Lett.~B315:139, 1993, hep-ph/9305258.
\bibitem{O} \verb$http://lepewwg.web.cern.ch/LEPEWWG/plots/summer2005/s05_stu_contours.eps$
\bibitem{W} B.~Holdom, Phys.~Rev.~Lett.~60:1233, 1988.
\bibitem{G} E.~Ma and J.~Pantaleone, Phys.~Rev.~D40:2172, 1989.
\bibitem{A} ATLAS Detector and Physics Performance Technical Design Report, CERN/LHCC/99-15(1999), sections 18.2 and 18.3;
E.~Arik, S.~Atag, Z.~Z.~Aydin, A.~Celikel,  Z.~Cicek, A.~K.~Ciftci, A.~Mailov, S.~Sultansoy, G.~Unel, Phys.~Rev.~D58:117701, 1998.
\bibitem{P} B.~Holdom, Phys.~Rev.~D24:1441, 1981; B.~Holdom, Phys.~Lett.~B150:301, 1985; K.~Yamawaki, M.~Bando, and K.-i.~Matumoto, Phys.~Rev.~Lett.~56:1335, 1986; T.~W.~Appelquist, D.~Karabali, and L.~C.~R.~Wijewardhana, Phys.~Rev.~Lett.~57:957, 1986; A.~Cohen and H.~Georgi, Nucl.~Phys.~B314:7, 1989; B.~Holdom, Phys.~Lett.~B213:365, 1988; B.~Holdom, Phys.~Rev.~Lett.~62:997, 1989.
\bibitem{Q} R.~Sundrum, Nucl.~Phys.~B395:60, 1993, hep-ph/9205203.
\bibitem{tmass} B.~Holdom, Phys.~Lett.~B336:85, 1994, hep-ph/9407244.
\bibitem{D} B.~Holdom, Phys.~Rev.~D54:1068, 1996, hep-ph/9512298; B.~Holdom, Prog.~Theor.~Phys.~Suppl.~123:71, 1996, hep-ph/9510249; B.~Holdom, Talk given at International Workshop on Perspectives of Strong Coupling Gauge Theories (SCGT 96), Nagoya, Japan, 13-16 Nov 1996, hep-ph/9702250; B.~Holdom, Phys.~Rev.~D57:357, 1998, hep-ph/9705231;  B.~Holdom, Talk given at the Workshop on Fermion Mass and CP violation, Hiroshima, Japan, March 1998, hep-ph/9804312.
\bibitem{N} W.~A.~Bardeen, C.~T.~Hill, and M.~Lindner, Phys.~Rev.~D41:1647, 1990.
\bibitem{V} B.~Holdom, Phys.~Lett.~B226: 137, 1989; T.~Appelquist, M.~Einhorn, T.~Takeuchi, and L.~C.~R.~Wijewardhana, Phys.~Lett.~B232:211, 1989.
\bibitem{M} R.~S.~Chivukula, S.~B.~Selipsky, and E.~H.~Simmons, Phys.~Rev.~Lett.~69:575, 1992.
\bibitem{J} G.~Cacciapaglia, C.~Csaki, C.~Grojean, J.~Terning, Phys.~Rev.~D71:035015, 2005, hep-ph/0409126.
\bibitem{CP} B.~Holdom, Phys.~Rev.~D61:011702, 2000, hep-ph/9907361.
\bibitem{H} A.~Arhrib, W.-S.~Hou, ``Flavor Changing Neutral Currents involving Heavy Quarks with Four Generations'', hep-ph/0602035.
\end{thebibliography}
\end{document}